\documentclass[12pt]{iopart}
\usepackage{amsfonts}

\usepackage{eurosym}

\begin{document}

\title[Turning on gravity with the Higgs mechanism]{Turning on gravity with the Higgs mechanism}
\author{Stephon Alexander}
\address{Brown University, Department of Physics, Providence, RI, 02912}
\author{John D. Barrow}
\address{DAMTP, Centre for Mathematical Sciences, Cambridge University, Wilberforce
Rd., Cambridge CB3 0WA, United Kingdom}
\author{Jo\~{a}o Magueijo}
\address{Theoretical Physics, Blackett Laboratory, Imperial College, London, SW7 2BZ,
United Kingdom}

\vspace{10pt}
\begin{indented}
\item[] 25 February 2016
\end{indented}

\begin{abstract}
We investigate how a Higgs mechanism could be responsible for the emergence
of gravity in extensions of Einstein theory, with a suitable low energy
limit. In this scenario, at high energies, symmetry restoration could
\textquotedblleft turn off\textquotedblright\ gravity, with dramatic
implications for cosmology and quantum gravity. The sense in which gravity
is muted depends on the details of the implementation. In the most extreme
case gravity's dynamical degrees of freedom would only be unleashed after
the Higgs field acquires a non-trivial vacuum expectation value, with
gravity reduced to a topological field theory in the symmetric phase. We
might also identify the Higgs and the Brans-Dicke fields in such a way that
in the unbroken phase Newton's constant vanishes, decoupling matter and
gravity. We discuss the broad implications of these scenarios.
\end{abstract}


\maketitle






\section{Introduction}

The Higgs mechanism is a central aspect of modern particle physics~\cite%
{higgs,EB,kibble}. At its core lies the idea that the vacuum and its
perturbative excitations do not partake of the full set of symmetries of the
underlying theory. This happens because the Higgs field acquires a
non-trivial vacuum expectation value (VEV), and fluctuations around it are
not as symmetrical as the Lagrangian of the theory. The symmetries thus
spontaneously broken are expected to be explicitly restored at high
energies, when the potential loses its 'Mexican hat' shape, allowing the
field to rest at the origin. The Higgs mechanism is responsible for giving
masses to gauge bosons (which must be fundamentally massless, due to the
gauge symmetry), as well as to fermions.

The possibility that the Higgs mechanism might have gravitational content is
intriguing, and has been considered before both in a theoretical framework,
with very different slants, all orthogonal to the approach in our Letter
(e.g.~\cite{Dehnen,Alexander1,percacci}), and from the observational point
of view, e.g. regarding spectral lines~\cite{Onofrio}. In this Letter we
consider a radical prospect in this regard: that gravity only properly
\textquotedblleft switches on\textquotedblright\ once the Higgs field
acquires a non-trivial VEV. This could occur in relation to the electroweak
Higgs doublet or with any other Higgs-like field, associated with any group
or pattern of symmetry breaking. In any such scenario, as we probe higher
and higher energies, the symmetries spontaneously broken by the Higgs field
are eventually restored, at which point gravity \textquotedblleft turns
off\textquotedblright , dropping out of the dynamical picture to some degree.

The precise sense in which this turn-off happens depends on the
technicalities of the implementation of this broad idea. For example, we may
adopt the view that Einstein gravity results from applying a constraint (the
so-called simplicity constraint) to a topological field theory (the
\textquotedblleft BF\textquotedblright\ theory introduced or reviewed in~%
\cite{Pleb,CJD,Speziale}, and defined in the next Section). The constraint
enters the action via a Lagrange multiplier, but suppose that the modulus of
the Higgs field multiplies this term. Then, in the symmetric phase (with $%
\phi _{0}=0$), the constraint would be turned off, rendering gravity a
topological field theory, with the dynamical degrees of freedom unleashed
only in the broken phase (with $\phi _{0}\neq 0$, turning on the
constraint). In this realization, at high energies there are no gravitons.
In addition the vacuum solutions are all of a topological nature. For
another discussion of the possible gravitational relevance of the Higgs
mechanism see ref \cite{per}, and for a different approach see ref \cite{rin}%
.

Another possibility follows from identifying the Higgs field (or any
Higgs-like field) with the Brans-Dicke field~\cite{BransDicke} or any
variation thereof~in a more general scalar-tensor gravity theory \cite{Fujii}%
, so that the Higgs VEV determines the strength of the gravitational
interaction, as fixed by the effective Newton's \textquotedblleft
constant\textquotedblright\ $G_{eff}$. By choosing a suitable coupling
function between the field and the Ricci scalar, we can ensure that $%
G_{eff}=0$ in the unbroken phase, but retain $G_{eff}\neq 0$ in the broken
phase. This effectively decouples matter and gravity in the unbroken phase,
so that there are still free gravitons at high energies, but there is no way
to produce them other than by vacuum quantum fluctuations. The
self-interaction of the graviton is also turned off. In this scenarios there
are still non-trivial vacuum solutions, just as with the usual scalar-tensor
theories; however singularities induced by matter are no longer present.

In our model we make a minimum of theoretical assumptions. In particular, we
will only assume the standard model  and general relativity, or similar
gravity theory. These provide gauge and general covariances. By combining
general covariance with the SU(2) gauge invariance, the Higgs field is
allowed to couple with general relativity with an SU(2) invariant (singlet
representation) of a functional of the Higgs field, $f(\phi )$~\footnote{The function $f(\phi)$ can in principle arise from string theory due to  the compactification (dimensional reduction) of the full 10 dimensional theory to our 4 dimensional world. String theory will yield a volume factor containing information of the extra six dimensions in the form of a function that could in principle have the required features of our function.  This calculation is beyond the scope of this letter and we shall pursue it in an upcoming work}. So, all
gauge invariant functions are allowed including Higgs multiplets in our
model, as in any gauge invariant extension to the standard model. The only
difference is that the gauge invariant functional couples to the Ricci
scalar. Even with a Higgs multiplet, it is important that the invariant
functional, $f(\phi ),$ vanishes in the symmetric phase. So, in general, if
one component of the multiplet is vanishing, it will not be sufficient just
to satisfy $f(\phi )=0$ to switch gravity off. In the Higgs mechanism, the
mass terms arise from the Higgs VEV, and this is where gravity turns off.
Our model assumes the minimal Higgs doublet, so there is only a VEV for the
Higgs. The various values of the Yukawa couplings, which control the masses
of the matter fields, do not play a role in switching gravity on or off---only the Higgs VEV controls that.

\section{The topological model}

Let us consider in more detail the first of these possibilities. Consider
the modified BF action~\cite{Pleb,CJD,Speziale}: 
\begin{equation}
S_{BF}=\int B^{IJ}\wedge F_{IJ}-\frac{1}{2}|\phi |\Phi _{IJKL}B^{IJ}\wedge
B^{KL},  \label{BF}
\end{equation}%
where $F^{IJ}(\omega )$ is the curvature of the connection $\omega ^{IJ}$
taking values in an algebra (assumed here to be $SO(3,1)$ or $SO(4)$, but it
could also be $SU(2)$); $B^{IJ}$ is a 2-form taking values in the same
algebra, and $\Phi _{IJKL}$ is a Lagrange multiplier enforcing the
simplicity constraint. Here, the indices run as $I=\{0,i\}=0,1,2,3$, $\mu
=0,1,2,3$. If $|\phi |$ is a non-vanishing constant it is known that this is
a constrained version of BF theory such that the action is equivalent to the
Cartan formulation of Einstein theory (i.e. a first-order formulation which
reduces to the usual second order in the absence of spinors; see~\cite%
{Cartan-review,Cartan-review2} for reviews). However, if $|\phi |=0$ (i.e.
if the simplicity constraint is not enforced) the theory is just plain BF
theory, i.e. it is a topological field theory. In both cases a densitized
metric can be constructed according to: 
\begin{equation}
\sqrt{-g}g_{\mu \nu }=\frac{1}{12}\epsilon _{ijk}\,\epsilon ^{\alpha \beta
\gamma \delta }B_{\mu \alpha }^{i}B_{\beta \gamma }^{j}B_{\delta \nu }^{k}.
\label{g}
\end{equation}%
%
%
%
and let us assume a convention so that a Lorentzian metric has signature $%
-+++$. The field $\phi $ can be a real scalar, or a complex doublet,
depending on how close to the standard model we want the model to be. All we
require is that $\phi $ be Higgs-like, so that its action is of the form: 
\begin{equation}
S_{Higgs}=\int d^{4}x\,\sqrt{-g}\left( -\frac{1}{2}(\partial _{\mu }\phi
)(\partial ^{\mu }\phi )-V(\phi )\right) 
\end{equation}%
where $V(\phi )$ is a 'Mexican hat' potential, which at high temperatures is
flattened into a simple bowl.

This theory has two types of solutions corresponding to two phases,
connected by a phase transition. At low temperatures, $V(\phi )$ has a
'Mexican hat' shape, so that the vacuum solutions lie on its rim, with the
non-trivial VEV $|\phi |=\phi _{0}\neq 0$. Even though there is a
topological sector~\cite{reinsen,freidel} in this phase, there is also a
dynamical sector, following from the solution: 
\begin{equation}
B^{IJ}=\frac{1}{2}\epsilon _{\;\;\;LM}^{IJ}e^{L}\wedge e^{M}
\end{equation}%
written in terms of $e^{I}$ (to be interpreted as the tetrad). For
non-degenerate tetrads, this can be inserted back into action (\ref{BF}),
yielding the Einstein-Cartan action. The torsion-free solutions then reduce
this action to the Einstein-Hilbert action. At high temperatures, however, $%
V(\phi )$ becomes a bowl potential, with a vacuum solution at $\phi =0,$ so
that the simplicity constraint cannot be imposed. Then, as is well known,
only the topological sector exists.

The implications for cosmology and quantum gravity are far-reaching. It
would appear that the Big Bang model at high temperatures gives way to a
simple topological solution~\cite{AlexanderCrane}. For isotropic cosmologies
where the spatial curvature is zero the cosmological singularity is
therefore removed\footnote{%
An anisotropic solution could still have a singularity, for example becoming
Kasnerlike as in the non-topological phase. We intend to pursue this issue
in a forthcoming work}. This is indeed the case in this model, where at high
energies the Higgs field switches off the graviton degrees of freedom. The
quantization of the ensuing topological field theory is standard \cite{Crane}%
.

A further layer can be uncovered by allowing the $\Phi _{IJKL}$ to have a
trace (a standard way to introduce a cosmological constant). Symmetry
dictates that we can have a renormalizable coupling of a Higgs-dependent
cosmological constant to gravity, $\Lambda _{0}\rightarrow \Lambda (\phi )$.
It turns out that this leads to even more interesting implications. The
action then becomes: 
\begin{equation}
S_{BF\Lambda }=S_{BF}+\Lambda (\phi )B^{IJ}\wedge B_{IJ}.
\end{equation}%
Here we identify the cosmological constant with the Higgs potential $\Lambda
(\phi )=V_{H}(\phi )=\lambda (\phi ^{2}-\eta ^{2})^{2}$. Upon variation with
respect to the Higgs we get the following condition 
\begin{equation}
D_{\mu }D^{\mu }\phi =\Phi _{IJKL}B^{IJ}\wedge B^{KL}+\Lambda (\phi
)^{\prime }B^{IJ}\wedge B_{IJ}
\end{equation}%
and the Einstein equations, 
\begin{equation}
F^{IJ}=B^{IJ}+\Lambda (\phi )^{\prime }B^{IJ}+L_{\phi }B^{IJ}.
\end{equation}%
Here $L_{\phi }$ is the Higgs field Lagrangian.  In the symmetric phase, $\phi =0$, and $\Lambda =\Lambda _{0}$, so that $F^{IJ}=B^{IJ}$, which yields the equations of topological gravity with a non-vanishing cosmological constant. These
theories have been dealt with in Lorentzian spin-foam quantization and also
in relation with the Kodama state \cite{Kodama,Smolin,Joao1,Joao2}.

Here, we see that the missing link regarding the issues of
non-normalizability in such theories may be the Higgs dependence of the
cosmological constant. The non-normalizability of topological quantum
gravity with a cosmological constant may be signifying an instability to
decay to a state of vanishing cosmological constant. For example,
semi-classically in the symmetry-breaking phase, $\phi =\phi _{min}$, the
cosmological constant vanishes and we get dynamical gravity. There is an
intermediate state where the tachyonic instability sets in and the Higgs is
rolling down to its minimum. Here we see that the cosmological constant is
dynamical and we have a situation where gravity is turned on and the
cosmological constant is varying with the evolution with the Higgs field.
The Higgs can tunnel from the false vacuum to the true vacuum and this may
serve as a potential way to explain why the true cosmological constant might
be zero, or at least much smaller than $\mathcal{O}(10^{-122})$, today. This
would require the dark energy responsible for the acceleration of the
universe to be supplied by a slowly varying energy field that has come to
dominate the universe at late times.  

\section{The scalar-tensor varying-$G$ model}

As a second, less radical realization for the proposal in this Letter, we
could seek to identify the Brans-Dicke field, or similar, and a Higgs-like
scalar field undergoing symmetry restoration at a phase transition~\cite%
{Dehnen}. However, such a generic model must be fundamentally modified
before it is put to service regarding the main idea in this paper:
\textquotedblleft switching off\textquotedblright\ gravity in the unbroken
phase. We seek a model in which $G_{eff}=0$ in the unbroken phase, when $%
\phi =0$. But in the simplest such model one has $G_{eff}\sim 1/\phi $, so
that $G_{eff}\rightarrow \infty $ in the unbroken phase, instead. We must
therefore look within the more general class of scalar-tensor theories, with
actions of the form: 
\begin{equation}
S=\int d^{4}x\,\sqrt{-g}\left( \frac{f(|\phi |)}{16\pi G_{0}}R-\frac{1}{2}%
(\partial _{\mu }\phi )(\partial ^{\mu }\phi )-V(\phi )\right) .  \label{S}
\end{equation}%
This is the action in the frame in which there is minimal coupling between
matter and gravity, pinning down a physical frame. For this model, we have~%
\cite{Tsujikawa}: 
\begin{equation}
G_{eff}=G_{0}\frac{1}{f}\frac{f+4f^{\prime 2}}{f+3f^{\prime 2}}\sim \frac{%
G_{0}}{f(|\phi |)}
\end{equation}%
so that 
\begin{equation}
G_{eff}\sim \frac{1}{f}\rightarrow 0
\end{equation}%
if $f\rightarrow \infty $ as $|\phi |\rightarrow 0$. When $\phi =\phi _{0}$
(the minimum of the 'Mexican hat' potential) we should have $f=1$. Any such
function realizes our idea. However turning off the coupling between matter
and gravity is quite different from turning off all of gravity's dynamic
degrees of freedom. Gravitons still exist. The complex array of vacuum
solutions known to exist in scalar tensor theories, both isotropic and
anisotropic, are still possibilities for the early universe.

These models are potentially very interesting for modelling the early
universe, and discussing various potential cosmogonic scenarios, as we now
illustrate. First, consider a crude cosmological model in general relativity
(GR) with an interval of time during which $G=0$ for $t_{1}<t<t_{2}$ and $%
G=G_{0}$ otherwise. This simple square-well gravity can be modelled by
multiplying the usual fluid energy-momentum tensor in Einstein's equations
by the sum of Heaviside functions 
\begin{equation}
Y_{12}=1-H(t-t_{1})+H(t-t_{2}),  \label{Y}
\end{equation}%
where $H(x)\geq 0$ for $x\geq 0$ so $Y_{12}=0$ for $t_{1}<t<t_{2}$ and zero
otherwise. The Friedmann equation is 
\begin{equation}
3\left( \frac{\dot{a}}{a}\right) ^{2}=8\pi G_{0}\rho Y_{12}-\frac{3k}{a^{2}}%
+\Lambda .  \label{fr}
\end{equation}%
For radiation (with $\rho \propto a^{-4}$) we have $a^{2}(t)\propto
Y_{12}t-kt^{2}$ $=-kt^{2}$ when $\Lambda =0$ for $t_{1}<t<t_{2}$ and so we
must have $k=-1$ for a solution to exist, and then $a(t)=t$ and there is a
coasting period of Milne expansion with no particle horizon present. When $%
\Lambda \neq 0$ we have de Sitter solutions for all $k$ so long as $%
a^{2}>3/\Lambda $ with $a(t)\propto \cosh (t\sqrt{\Lambda /3})$ for $%
k=+1,a(t)\propto \exp (t\sqrt{\Lambda /3})$ for $k=0$, and $a(t)\propto
\sinh (t\sqrt{\Lambda /3})$ for $k=-1$. However, during the interval $%
(t_{1},t_{2})$ there can still be solutions even when $\Lambda =0$ and $%
k\geq 0$ because in reality there will always be departures from exact
isotropy which contribute (possibly very small) shear terms on the RHS of
eq.(\ref{fr}) proportional to $a^{-6}$ which drive the solution towards $%
a\propto t^{1/3}$ when $k=0$, for example.

Again, we see that turning-off $G$ still leaves some gravitational degrees
of freedom: it reduces the free data required to specify a general GR
cosmological solution on a spacelike hypersurface from 8 arbitrary functions
to 4 for a perfect fluid, and from 6 to 4 for a self-interacting scalar
field \cite{JDB}. If there is inhomogeneity, with $t_{1}(\vec{x})$ and $%
t_{2}(\vec{x}),$ then $Y_{12}(\vec{x})$ and the duration of this finite
period of zero-$G$ evolution, will be spatially varying and create residual
density variations at $t_{2}(\vec{x})$ even if $\rho $ were uniform before $%
t_{1}$.

This type of ``on-off-on'' gravity scenario could be implemented in the
action (\ref{S}) in a theory with an scalar potential of the form $%
f(\left\vert \phi \right\vert )\propto G_{eff}\propto Y_{12}^{-1}$. In this
case, we would not simply recover the vacuum solutions on the interval $%
(t_{1},t_{2}),$ as there are many vacuum solutions in generalized
scalar-tensor theories like Eq.~(\ref{S}) with non-constant $\phi $, just as
there are in the simplest case of Brans-Dicke gravity. The square-well form (%
\ref{Y}) can now be smoothed to the form $\tanh ^{2n}(\lambda \phi ),$ with $%
n\in 
\mathbb{Z}
^{+}$, to mimic the turn-off effect in $(t_{1},t_{2})$ during the very early
universe. We will examine the\ interesting consequences of such scenarios
elsewhere, but the general picture is one we have already sketched: one in
which the early universe is simpler during a finite interval of time, due to
the partial switching off of gravity.

\section{Conclusions}

To conclude, we thus see that a number of scenarios may be contemplated for
both the early universe and the puzzle of quantum gravity, depending on how
we implement our broad idea that the Higgs mechanism is responsible for
switching on gravity. In whatever realization, the Big Bang universe's early
stages (and possibly the Big Bang singularity itself) are replaced by a
different type of solution. In the first realization proposed here a
topological solution would model the early universe; in the second a vacuum
solution of a scalar-tensor theory would rule it. In both cases the
cosmogony is dramatically modified, with opportunities to remove the
cosmological singularity. The possibility that a similar phenomenon might
occur in the future would also modify the eschatology, rendering the future
of the Universe simpler. We stress that the phenomenon envisaged here is
different from that proposed in~\cite{Grf,dsr,dsr1}, where the muting of
gravity applies only to the cosmological perturbations. However the two
ideas could be complementary.

One further scenario we wish to pursue in the future, is based on the conjecture
that, as a converse to the expectation that all symmetries are restored at
high enough energies, perhaps they are all broken at low enough energies.
This has been discussed in connection with the breaking of $U(1)$ symmetry
at low energy, thus creating a photon mass and even the possibility of
photon oscillations via $U(1)\times U^{\prime }(1)$ (see~\cite{glash,BB}).
If such symmetry breaking in the future drove $G\rightarrow 0,$ then $t_{1}$
might exceed the present age and $t_{2}=\infty$. The future cosmological
evolution of unbound structures at $t>t_{1}$ would not be significantly
altered, since it is already destined to be dominated by $G$-independent
factors, like curvature, frozen-in anisotropies with constant shear to
Hubble ratio, or $\Lambda $, in the future of open universes (cf. \ref{fr}%
)). However, all bound structures would unbind and black hole horizons would
shrink to zero. In this sense the future would be asymptotically simpler.

Regarding the conundrum of quantum gravity, we recall that most of the work
attempting to solve the puzzle has been performed in the absence of matter
fields. It has been speculated that the problems with the UV limit of the
theory could be resolved with the addition of matter. This could indeed be
the case in the scenarios considered here: specifically the Higgs field
could turn off gravity at high energies, either trivializing its UV limit by
rendering it topological (as in~\cite{Krasnov}), or switching off its
interactions and self-interactions. This would be the ultimate asymptotic
freedom. The full implications of this scenario should be studied in more
detail, but it should be clear that this is a new avenue of enquiry.


We close with two interesting avenues of further study for this type of
models. As with any phase transition with spontaneous symmetry breaking we
know that a cosmic network of topological defects may be produced, with a
morphology dependent on the homotopy groups of the quotient of the full
group and the broken subgroup. For example, for a real scalar Higgs field
one would have domain walls. Inside the topological defect the symmetry
remains unbroken, even after the transition, the field remaining stuck on
the false vacuum. The implication, in our scenario, is that gravity would
remain switched off at the core of any defect associated the symmetry
breaking responsible for turning on gravity. Such solutions could be very
interesting, both theoretically and phenomenologically. On a different front
the weak field limit of these theories is also bound to contain interesting
phenomenology. This was already discussed in the past in~\cite{Dehnen}.
Corrections to the Newtonian potential will be more involved than just a
shift in $G$ or of the Yukawa type. We defer to a future publication the
full investigation of the ensuing constraints.

\section*{Acknowledgements}

JDB is supported by STFC and also acknowledges support and hospitality from
the Big Questions Institute, the School of Physics, and Faculty of Science
at the University of New South Wales, Sydney during this work. JM
acknowledges support from John Templeton Foundation, a STFC consolidated
grant and the Leverhulme Trust.


\section*{Bibliography}

\begin{thebibliography}{99}
\bibitem{higgs} P. Higgs (1964), 
Phys. Lett. \textbf{12} (1964) \textbf{132}; 
Phys. Rev. Lett. \textbf{13} (1964) 508.

\bibitem{EB} F. Englert and R. Brout, 
Phys. Rev. Lett. \textbf{13} (1964) 321.

\bibitem{kibble} G.S. Guralnik, C.R. Hagen and T.W.B. Kibble, 
Phys. Rev. Lett. \textbf{13} (1964) 585.

\bibitem{Dehnen} H. Dehnen, H.~Frommert, and F.~Ghabouss, Int. J Th. Phys., 
\textbf{31} (1992) 109.


\bibitem{Alexander1} S.~Alexander, A.~Marciano and L.~Smolin, 
Phys.\ Rev.\ D \textbf{89}, no. 6, 065017 (2014) 
[arXiv:1212.5246 [hep-th]]. 

\bibitem{percacci} R.~Percacci,  
PoS ISFTG , 011 (2009)  [arXiv:0910.5167 [hep-th]];  
R.~Percacci,  
Nucl.\ Phys.\ B \textbf{353}, 271 (1991)  doi:10.1016/0550-3213(91)90510-5 
[arXiv:0712.3545 [hep-th]].  

\bibitem{Onofrio} R.~Onofrio, 
Phys.\ Rev.\ D \textbf{82}, 065008 (2010)

\bibitem{Pleb} M. J. Plebanski, 
J. Math. Phys. \textbf{18} (1977), 2511.

\bibitem{CJD} R. Capovilla et al, 
Class. Quant. Grav. \textbf{8} (1991) 41.

\bibitem{Cartan-review} F.W. Hehl, P. Von Der Heyde, G.D. Kerlick, and J.M.
Nester 
Rev.Mod.Phys. , 48:393â\euro ``416, 1976.

\bibitem{Cartan-review2} M. Gockeler and T. Schucker, "Differential
geometry, gauge theories, and gravity", CUP, Cambridge 1989.

\bibitem{AlexanderCrane} S.~Alexander, L.~Crane and M.~D.~Sheppeard, 
gr-qc/0306079. 

\bibitem{Kodama} H.~Kodama, 
Phys.\ Rev.\ D \textbf{42}, 2548 (1990). 

\bibitem{Smolin} L.~Smolin, 
hep-th/0209079. 

\bibitem{Speziale} S.~Speziale, 
Phys.\ Rev.\ D \textbf{82}, 064003 (2010)


\bibitem{per} R. Percacci, PoS ISFTG2009:011 (2009) [arXiv:0910.5167]

\bibitem{rin} M. Rinaldi and L. Vanzo, [ arXiv:1512.07186]

\bibitem{Joao1} J.~Magueijo and D.~M.~T.~Benincasa, 
Phys.\ Rev.\ Lett.\ \textbf{106}, 121302 (2011) 
[arXiv:1010.3552 [gr-qc]]. 


\bibitem{Joao2} L.~Bethke and J.~Magueijo, 
Phys.\ Rev.\ D \textbf{84}, 024014 (2011) 
[arXiv:1104.1800 [gr-qc]]. 

\bibitem{reinsen} M. P. Reisenberger, Nucl. Phys. B \textbf{457}, 643 (1995)

\bibitem{freidel} R. De Pietri and L. Freidel, Class. Quant. Grav. \textbf{16%
} (1999) 2187

\bibitem{BransDicke} C.~Brans and R.~H.~Dicke, 
Phys.\ Rev.\ \textbf{124}, 925 (1961). 

\bibitem{Fujii} Y.~Fujii and K.~Maeda, \textit{The scalar-tensor theory of
gravitation}, CUP, Cambridge, 2003. 

\bibitem{Tsujikawa} S.~Tsujikawa, 
Phys.\ Rev.\ D \textbf{76}, 023514 (2007) 

\bibitem{JDB} J.D. Barrow, Phys. Rev. D \textbf{89}, 064022 (2014)

\bibitem{glash} H. Georgi, P. Ginsparg, and S.L. Glashow Nature \textbf{306}%
, 765 (1983)


\bibitem{Crane} J.~W.~Barrett and L.~Crane, 
Class.\ Quant.\ Grav.\ \textbf{17}, 3101 (2000) 
[gr-qc/9904025]. 

\bibitem{okun} L. Okun, Sov. Phys. JETP \textbf{56}, 502 (1982)

\bibitem{BB} J.D. Barrow and R. Burman, Nature, \textbf{307}, 14 (1984)

\bibitem{Grf} G.~Amelino-Camelia, M.~Arzano, G.~Gubitosi and J.~Magueijo, 
Int.\ J.\ Mod.\ Phys.\ D \textbf{24}, 1543002 (2015)

\bibitem{dsr} G.~Amelino-Camelia, M.~Arzano, G.~Gubitosi and J.~Magueijo, 
Phys.\ Rev.\ D \textbf{87}, no. 12, 123532 (2013) 
[arXiv:1305.3153 [gr-qc]].

\bibitem{dsr1} G.~Amelino-Camelia, M.~Arzano, G.~Gubitosi and J.~Magueijo, 
Phys.\ Rev.\ D \textbf{88}, no. 4, 041303 (2013) 
[arXiv:1307.0745 [gr-qc]].

\bibitem{Krasnov} K.~Krasnov, 
Proc.\ Roy.\ Soc.\ A \textbf{468}, 2129 (2012) 
[arXiv:1202.6183 [gr-qc]].
\end{thebibliography}
\end{document}